\documentstyle[prl,tighten,aps,epsfig,multicol,amsfonts, times]{revtex}

\begin{document}
\author{Daniel Jonathan${}^{1}$ and Martin B.
Plenio${}^{2}$}
\title{Light-shift-induced quantum gates for ions in thermal motion}
\address{ ${}^{1}$  DAMTP, Centre for Mathematical Sciences, University of Cambridge,
Wilberforce Road, Cambridge CB3 0WA, U.K. \\ ${}^{2}$ QOLS,
Blackett Laboratory, Imperial College, London SW7 2BW, U.K.}
\date{\today}
\maketitle

\begin{abstract}

An effective interaction between trapped ions in thermal motion
can be generated by illuminating them simultaneously with a single
laser resonant with the ionic carrier frequency. The ac
Stark-shift induces simultaneous `virtual' two-phonon transitions
via several motional modes. Within a certain laser intensity range
these transitions can interfere constructively, resulting in a
relatively fast, heating-resistant two-qubit logic gate.
\end{abstract}

\begin{multicols}{2}

Over the past few years, the implementation of a practical quantum
information processor has become a major goal for experimentalists
across a wide range of disciplines \cite{FortPhys}. Among the many
candidate scenarios for attaining this goal, one of the most
promising and well-known is the system of laser-cooled trapped
ions, first suggested by Cirac and Zoller (CZ) \cite{CZ,reviews}.
Several key features of their seminal proposal have already been
experimentally demonstrated \cite{exp_papers}. Meanwhile, the
inherent difficulty of these experiments has stimulated the
development of an array of alternative gate schemes
\cite{theor_papers,Sorensen99a,Sorensen00a,Jonathan00a} making
increasingly ingenious use of the ion trap's physics.

Here we combine the best features of two of these proposals with an
extra twist, constructing a two-qubit gate mechanism which is at once
(i) robust against environmental heating, (ii) relatively fast, and
(iii) requires relatively few experimental resources, namely a single
pulse by a single laser and no ground-state-cooling.
Our first ingredient is the idea, due to S\o rensen and M\o lmer
\cite{Sorensen99a}, of coupling different ions using Raman-like
two-phonon exchanges via a ``data bus" motional mode. Since the
mode is only excited `virtually', this allows the realisation of
gates that are relatively insensitive to the ionic vibrational
state, and can be implemented even in the presence of moderate
motional heating \cite{Sorensen99a}. This method's largest
drawback \cite{James00a} is the very low switching rate of the
gates, which are substantially slower than those obtained with
CZ's method. This is undesireable, since it increases the quantum
register's vulnerability to other sources of decoherence such as
technical noise in the experimental apparatus. The problem has
been tackled to an extent in ref. \cite{Sorensen00a}, where an
elegant means of considerably increasing the gate speed is
provided. However, this solution requires heavily entangling the
internal and motional variables during the gate operation, making
it more sensitive to heating.

Our second ingredient is a method we have recently proposed for
obtaining relatively fast gates between ground-state-cooled ions
\cite{Jonathan00a}. The method relies on the ac Stark-shift
(light-shift) induced by driving an ion with a resonant laser.
Normally, this driving merely leads to Rabi flopping between the
ionic states, realising a one-qubit gate. However, when the
induced level-splitting is equivalent to exactly one motional
energy quantum, the influence of off-resonant motion-affecting
transitions can gradually build up, leading to the exchange of
excitations between the ion's internal and vibrational states.
This can then be used to derive a fast CZ-like two-qubit gate
between any two ions in a linear chain. As with CZ's method,
however, this scheme requires ground-state cooling and is highly
sensitive to motional heating.

In what follows we show a way of combining these ideas and their
resulting benefits. Specifically, we propose
to use the light-shift effect to drive `virtual' two-phonon
transitions via several motional modes {\em simultaneously}. Each
mode functions as a parallel ``data bus", with an overall gate
resulting from the interference of various `bus' paths. The gates
have speeds intermediate between those in refs. \cite{Sorensen99a}
and \cite{Sorensen00a}, but remain insensitive to heating at all
times. Moreover, only a single laser beam is needed, instead of
the bichromatic illumination used in
\cite{Sorensen99a,Sorensen00a}. As in \cite{Jonathan00a}, the same
beam can also be used to drive single-qubit gates.
\begin{center}
\begin{figure}[hbt]
\leavevmode \epsfysize=3.5cm \epsfbox{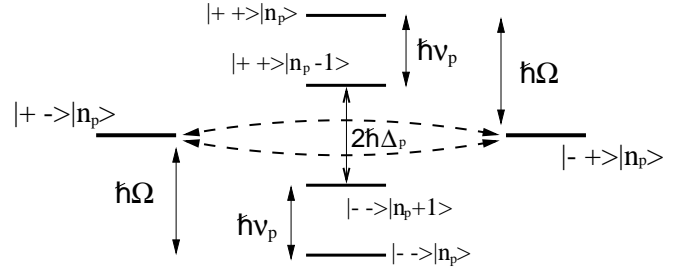}
\begin{minipage}{3.4truein}
\smallskip \smallskip \caption{A laser resonant with the carrier
transition illuminates two ions, splitting each dressed level pair
$\left| \pm \right\rangle $ by $ \hbar\Omega$. When this splitting
is detuned from all vibrational frequencies $\nu_q$, states $
\left| +- \right\rangle \left| \left\{n_q\right\} \right\rangle$
and $\left| -+ \right\rangle \left| \left\{n_q\right\}
\right\rangle$ are degenerate with each other and non-degenerate
with all other levels (for clarity, only levels of a single mode
$p$ are shown). Coherent oscillations are then induced between
them, at a frequency independent of $\left\{n_q\right\}$. This is
due to the existence, for each mode, of `virtual' two-phonon
transition paths connecting the two states, via levels $\left| ++
\right\rangle \left| n_p-1 \right\rangle$ and $\left| --
\right\rangle \left| n_p+1 \right\rangle$.} \label{hotshiftfig}
\end{minipage}
\end{figure}
\end{center}

Let us begin with an intuitive description of our essential idea
(Fig. \ref{hotshiftfig}). Consider a chain of $N$ ions, two of
which are simultaneously and equally illuminated by a laser
resonant with the carrier frequency. At first sight, this simply
causes these ions' internal states to flop periodically at the
Rabi frequency $\Omega$. Equivalently, each ion suffers a
level-splitting of its (semiclassical) dressed states $\left| \pm
\right\rangle = 1/\sqrt{2}\left( \left| g\right\rangle \pm \left|
e\right\rangle \right)$ by $\hbar\Omega$. The presence of the
trap, however, opens the possibility of {\em transitions} between
these dressed levels, via the absorption and emission of phonons
\cite{Jonathan00a}. Which transitions are energetically allowed
depends on the magnitude of the level-splitting. For example, when
$\Omega$ is exactly equal to the frequency $\nu_{p}$ of one of the
ions' collective motional modes, then the manifolds $\left\{
\left| --\right\rangle \left| n_p+1\right\rangle,\left|
-+\right\rangle \left| n_p\right\rangle,\left| +-\right\rangle
\left| n_p\right\rangle,\left| ++\right\rangle \left|
n_p-1\right\rangle\right\} $ are degenerate and one expects
transitions altering the phonon number in that mode. This is
interesting, but inadequate for obtaining a
motion-independent ion-ion gate. If, however, $\Omega $ is slightly {\it
detuned } from this value, then transitions
involving changes in $n_p$  become off-resonant and are
suppressed. One is left only with 
oscillations between $\left| -+\right\rangle \left|
n_p\right\rangle$ and $\left| +-\right\rangle \left|
n_p\right\rangle$, which can be interpreted as occurring via the
Raman-like absorption and emission of `virtual' phonons.

Now, just as in ref. \cite{Sorensen99a}, each of these virtual
transitions can occur via two different paths: one `via' level
$\left| ++\right\rangle \left| n_p-1\right\rangle$, with amplitude
proportional to $n_p$ and positive detuning $\Delta_p$, and one
`via' level $\left| --\right\rangle \left| n_p+1\right\rangle$,
with amplitude proportional to $n_p+1$ and negative detuning $
-\Delta_p $. The overall transition amplitude will thus be the
{\em difference} of these terms; remarkably, the $n_p$-dependence
should cancel out, and we can expect {\em
motion-state-independent} transitions between the internal states
$\left| -+\right\rangle $ and $\left| +-\right\rangle $! Of
course, the ion system contains in fact several motional modes,
each of which will lead  to a separate resonance in the Rabi
frequency. However, if $\Omega $ is sufficiently detuned from all
these resonances, the argument above holds {\em simultaneously}
for all modes. In other words, the overall transition amplitude
between $\left| -+\right\rangle $ and $\left| +-\right\rangle $
should result from the interference of multiple Raman-like paths,
two for each separate mode. Thus, the totality of modes are used
together
as  a `collective data bus' connecting the two ions.

To see that all this is really the case, consider the Hamiltonian
of this system. Within the Lamb-Dicke  limit
\begin{equation}
\eta _{p}\sqrt{(\bar{n}_{p}+1)}\ll 1  \label{LDlim}
\end{equation}
(where $\eta _{p}$ is the Lamb-Dicke (LD) parameter of the
$p^{th}$ collective mode of the ions and $\bar{n}_{p}$ is the
average number of phonons in that mode), this can be written as
\cite{reviews}
\begin{eqnarray}
    H \simeq \frac{\Omega}{2} \sum_{j=1}^{2}
     e^{i\phi _{j}}\sigma_{j+}
    \left[ 1+\sum_{p=1}^{N}i\eta _{jp}\left[ a_{p}^{\dagger}e^{i
    \nu_{p}t}+ h.c.\right] \right] \! + h.c. \label{Ham}
\end{eqnarray}
Here $\Omega $ is the effective Rabi frequency of the laser-ion
interaction (assumed to be equal for both ions), $\phi_{j}$ is the
laser's phase at the position of the $j^{th}$ ion, $\nu_{p}$ is
the frequency of the $p^{th}$ collective mode ($p=1$ for the
centre-of-mass (CM) mode, $p=2$ for the `breathing' mode, etc.)
and $\eta_{jp}$ is an `effective' LD parameter incorporating the
relative displacement of the $j^{th}$ ion in the $p^{th}$ mode
\cite{reviews,Note1}. We have also set $\hbar=1$. In what follows
we will assume, with no loss of generality, $\phi _{1}=\phi
_{2}=0$ \cite{Note4}.

Following now a reasoning analogous to that of ref. \cite{Jonathan00a}, we
find that resonance conditions arise for particular values of $\Omega $.
This can be most easily seen by transforming into a dressed-state picture
defined by the operator
\begin{equation}
V\left( t\right) \equiv \exp \left( \frac{i\Omega t\sigma _{1z}}{2}
\right) R_{1}\otimes \exp \left( \frac{i\Omega t\sigma
_{2z}}{2}
\right) R_{2}  \label{drespic}
\end{equation}
where $R_{j}=\frac{1}{\sqrt{2}}{{1\,1 \choose -1\,1} }_{j}$ and in
our convention $\left| e\right\rangle = {1 \choose 0}$ . The
Hamiltonian becomes then
\begin{equation}
H^{\prime }=\frac{\Omega}{2} \sum_{p=1}^{N}iJ_{p+}^{\prime }\left[
e^{i\Delta _{p}t}a_{p}+e^{i\gamma _{p}t}a_{p}^{\dag }\right]
+h.c.\;,  \label{dresHam}
\end{equation}
where we define $J_{p\pm }^{\prime }\equiv \sum_{j=1}^{2}\eta
_{jp}\sigma _{j\pm }^{\prime }$ ; $\Delta _{p}\equiv \Omega -\nu
_{p}$, $\gamma _{p}\equiv \Omega +\nu _{p}$ and primes indicate
operators defined in the dressed-state picture. One can see that,
when $\Delta _{p}=0$, the dynamics is dominated by terms of the
form $\sigma _{j+}^{\prime }a_{p}$, which induce resonant
collective excitations of the ions accompanied by the exchange of
motional quanta.

In what follows we are interested in the complementary regime far from these
resonance conditions. Specifically, let us consider the limit where the
detunings $\Delta _{p}$ of the Rabi frequency are large with respect to the
resulting secular frequency of the time evolution, i.e.
\begin{equation}
\Delta _{p}\gg \eta_{jp}\Omega = \eta _{jp}\left( \nu _{p}+\Delta
_{p}\right). \label{deltalimit}
\end{equation}
In this case, standard time-averaging arguments (see e.g. the
Appendix in \cite{James00a}) can be used to obtain a
time-independent {\it effective }Hamiltonian, given by
\begin{equation}\label{JamesHam}
H_{eff}^{\prime }\approx \frac{\Omega ^{2}}{4}\sum_{p=1}^{N}\left[
\frac{\left[ J_{p-}^{\prime }a_{p}^{\dagger },J_{p+}^{\prime
}a_{p}\right]}{\Delta _{p}} +\frac{\left[ J_{p-}^{\prime
}a_{p},J_{p+}^{\prime }a_{p}^{\dagger }\right]}{\gamma _{p}}
\right]
\end{equation}

Evaluating the commutators, we get after some algebra $\left[ J_{p-}^{\prime
}a_{p}^{\dagger },J_{p+}^{\prime }a_{p}\right] =B_{p}^{\prime
}-A_{p}^{\prime }+k_{1}$; $\left[ J_{p-}^{\prime }a_{p},J_{p+}^{\prime
}a_{p}^{\dagger }\right] =B_{p}^{\prime }+A_{p}^{\prime }+k_{2}$, where
\begin{mathletters}
\begin{eqnarray}
A_{p}^{\prime } &=&\eta _{1p}\eta _{2p}\left[ \left| e^{\prime }g^{\prime
}\right\rangle \left\langle g^{\prime }e^{\prime }\right| +\left| g^{\prime
}e^{\prime }\right\rangle \left\langle e^{\prime }g^{\prime }\right| \right]
\\
B_{p}^{\prime } &=&\left( n_{p}+\frac{1}{2}\right) \left[ \left(
\eta _{1p}^{2}+\eta _{2p}^{2}\right) \left[ \left| g^{\prime
}g^{\prime }\right\rangle \left\langle g^{\prime }g^{\prime
}\right| -\left| e^{\prime }e^{\prime }\right\rangle \left\langle
e^{\prime }e^{\prime }\right| \right] \right.   \nonumber \\
&+&\left. \left( \eta _{1p}^{2}-\eta _{2p}^{2}\right) \left[
\left| g^{\prime }e^{\prime }\right\rangle \left\langle g^{\prime
}e^{\prime }\right| -\left| e^{\prime }g^{\prime }\right\rangle
\left\langle e^{\prime }g^{\prime }\right| \right] \right]
\label{Bp'}
\end{eqnarray}
\end{mathletters}
and where $k_{1},k_{2}$ are constants that may be disregarded by a
suitable redefinition of the zero of energy. Thus
\begin{eqnarray}
H_{eff}^{\prime } &\approx &
-\omega \left[ \left| e^{\prime }g^{\prime }\right\rangle
\left\langle g^{\prime }e^{\prime }\right| +\left| g^{\prime
}e^{\prime }\right\rangle \left\langle e^{\prime }g^{\prime
}\right| \right] +\sum_{p=1}^{N}\frac{\Omega ^{3}B_{p}}{2(\Omega
^{2}-\nu _{p}^{2})} \label{effHam}
\end{eqnarray}
where
\begin{equation}
\omega =\frac{\Omega ^{2}}{2}\sum_{p=1}^{N}\frac{\eta _{1p}\eta _{2p}\nu _{p}}{%
\Omega ^{2}-\nu _{p}^{2}}.  \label{gatefreq}
\end{equation}
Of course, this Hamiltonian holds only in the dressed picture
defined in Eq.(\ref{drespic}). In the `standard' picture (i.e.,
the one corresponding to Eq. (\ref{Ham})), the time evolution
operator is $U(t)\equiv V^{\dag}(t)\exp(-itH_{eff}^{\prime})V(0)$.

In order to analyse this result, let us first consider the
case $N=2$, where
 $\eta_{11}=\eta_{21}= \sqrt[4]{3}\eta_{12}= -\sqrt[4]{3}\eta_{22}$
\cite{reviews}. The second term in Eq. (\ref{Bp'}) then vanishes,
and $H_{eff}^{\prime }$ splits into one term coupling internal
states $\left|g^{\prime}e^{\prime}\right\rangle$ and
$\left|e^{\prime}g^{\prime}\right\rangle$, and another affecting
only $\left|g^{\prime}g^{\prime}\right\rangle$ and
$\left|e^{\prime}e^{\prime}\right\rangle$. Translating back into
the ``standard" picture, and including also the motional
variables, the first term leads to transitions between each pair
of states $\left| +-n_{1}n_{2}\right\rangle ,\left|
-+n_{1}n_{2}\right\rangle $, as expected from Fig.
\ref{hotshiftfig} (note that all time-dependent phases originating
in $V\left( t\right) $ cancel out). Furthermore, as was also
previously suggested, the frequency $\omega$ of the transitions is
independent of the phonon numbers $\left\{n_{p}\right\}$. Finally,
it is clear from Eqs. (\ref{effHam}),(\ref {gatefreq}) that
this coupling results from the interference of contributions from
both modes. The interference is constructive (both terms
contribute positively to $\omega$, leading to faster oscillations)
when $\nu_1\leq \Omega \leq \nu_2 =\sqrt{3} \nu_1$. Note also that
in general the further $\Omega$ is to $\nu_q$, the smaller is the
amplitude in $\omega$ of the transition path via the corresponding
mode. This effect might be useful to screen out `bad' modes with
relatively high decoherence rates, such as the centre-of-mass mode
in multi-ion traps \cite{James98b}.

\begin{center}
\begin{figure}[hbt]
\epsfig{file=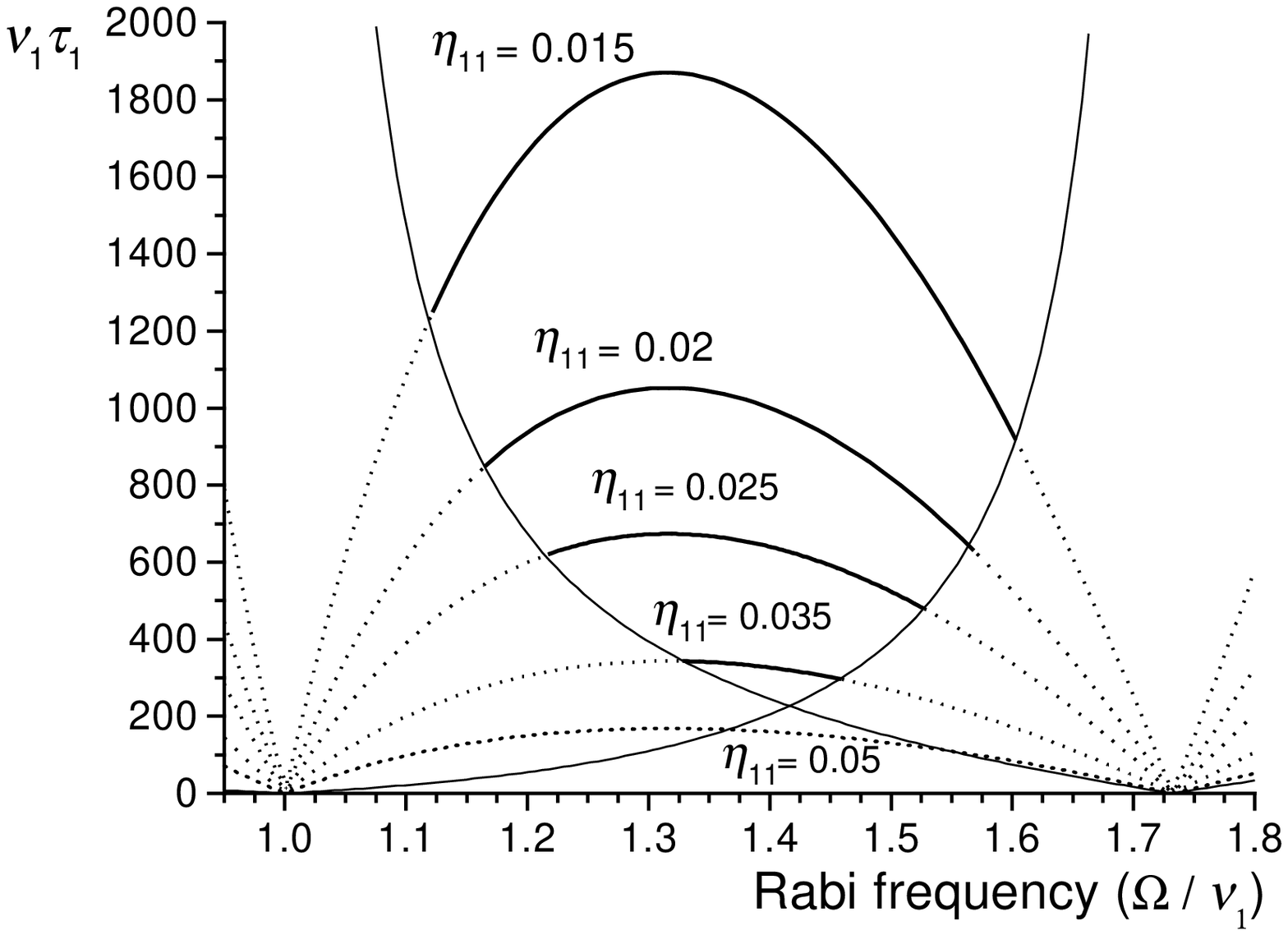, bbllx=31, bblly=400, bburx=514, bbury=760,
width=6.9cm}
\begin{minipage}{3.4truein}
\smallskip \smallskip \caption{Time required for the creation of
a Bell state starting from the initial state $\left|
+-\right\rangle$, calculated from Eq. (\ref{gatefreq}). Thin solid
curves represent the limits imposed by  Eq. (\ref{deltalimit}).
The method should hold well in the region above both curves.}
\label{fig:speed}
\end{minipage}
\end{figure}
\end{center}
Meanwhile, states $ \left| ++n_{1}n_{2}\right\rangle $ and $\left|
--n_{1}n_{2}\right\rangle $  acquire time-dependent phases,
originating both from $V\left( t\right)$ and from
$H_{eff}^{\prime}$. The latter do depend on the phonon numbers and
can lead to unwanted correlations between internal and motional
variables, and thus to decoherence of the internal state. These
results are analogous to those obtained in ref. \cite{Sorensen99a}
for the case where each ion is separately illuminated by a
different laser, one slightly detuned from the first red sideband
frequency and the other from the first blue one. As in that case,
here it turns out that the motion-state-dependent component of the
evolution can be cancelled by altering the Hamiltonian in the
course of the gate operation itself (see below).

As a result, maximally entangled states of the two ions can be
produced after a time $\tau_1 = |\pi/4\omega|$. To see this, note
that the evolution operator $U(\tau_1)$ is essentially a
`$\sqrt{SWAP}$' gate with respect to the $\left| \pm\right\rangle$
basis: it turns the disentangled states $\left| +-\right\rangle
,\left| -+\right\rangle $ into the maximally entangled (or `Bell')
states $\left|\beta_{\pm}\right\rangle \equiv \left( \left|
+-\right\rangle \pm i\left| -+\right\rangle \right)/\sqrt{2} $,
while leaving $\left|++\right\rangle$ and $\left| --\right\rangle$
unaffected (up to motion-state-independent phases
$\phi_{\pm}(\tau_1)$ due to $V^{\dag}(t)$). We stress again that
this evolution occurs {\it regardless} of the initial motional
state, and even regardless of whether it might be changing (due to
heating effects) during the gate operation itself. Note further
that, after a pulse of duration $2\tau_1$, $\left| \pm
\mp\right\rangle \rightarrow i\left|\mp \pm\right\rangle$, $\left|
\pm \pm \right\rangle \rightarrow \exp(\pm i
\Omega\tau_1)\left|\pm \pm\right\rangle$, and so all four standard
basis states $\left| g(e)g(e)\right\rangle $ become maximally
entangled. In other words, a gate locally equivalent to a CNOT is
realised. Our method therefore allows the implementation of
universal quantum logic on the ion chain, even in the presence of
heating. The speed with which this is accomplished can be seen in
Fig. \ref{fig:speed}, where we use Eq. (\ref{gatefreq}) to plot
$\tau_1$ as a function of $\Omega$, $\eta_{11}$. The thin solid
lines indicate the validity limits imposed by choosing a factor of
10 in Eq. (\ref{deltalimit}). We can conclude that the method
should allow $\tau_1$ to be as low as a few hundred trap periods,
a performance intermediate between that in \cite{Sorensen99a} and
that of the enhanced but heating-sensitive method in
\cite{Sorensen00a}. A significant difference with regard to these
proposals is that our method should work in a different parameter
range, including a comparatively high Rabi frequency and a
comparatively small LD parameter. This may give it a substantial
speed advantage in situations where $\eta_{11}$ is forcibly small,
such as a tight trap containing many ions.

The trick which cancels the motion-state-dependent part of the
evolution is to use a `photon-echo'-like procedure
\cite{Sorensen99a}, whereby midway along the time evolution one
inverts the sign of the motion-dependent term in the Hamiltonian
(in our case $B_{p}$ in Eq.(\ref{effHam})). As a result, phases
acquired in the first half of the evolution are cancelled out
exactly during the second one. In the present setup, this idea can
be implemented in an experimentally simple way, by suddenly
shifting the laser phase by $\pi $ (using, for instance, an
electro-optic modulator). To see this, let us return to
eq.(\ref{Ham}). Shifting $\phi _{j}\rightarrow\phi _{j}+\pi $ is
equivalent to changing the sign of $\Omega $, which corresponds to
{\it exchanging the signs} of the light-shifts suffered by $\left|
+\right\rangle $ and $\left| -\right\rangle $). A development
analogous to eqs. (\ref{drespic})-(\ref{gatefreq}) results then in
an effective Hamiltonian identical to $H_{eff}^{\prime}$ but with
$\left| e^{\prime}\right\rangle $ and $\left|
g^{\prime}\right\rangle $ everywhere interchanged. It can be
easily seen that this leaves the motion-independent term $A_{p}$
intact, while changing the sign of $B_{p}$ as required. In fact,
following again ref. \cite{Sorensen99a}, in order to suppress
motional heating it is desireable to perform the sign inversion
several times during the course of the system's evolution
(specifically, at frequency $F=M/\tau,$where $\tau$ is the total
time during which the laser is applied and $M\gg1$ is an integer).
If $F\gg\omega,\Gamma$, where $\Gamma$ is the typical heating
rate, then the effects of the motion-dependent terms are cancelled
before heating can affect them \cite{Sornote}.

\begin{center}
\begin{figure}[hbt]
\leavevmode \epsfxsize=8cm \epsfbox{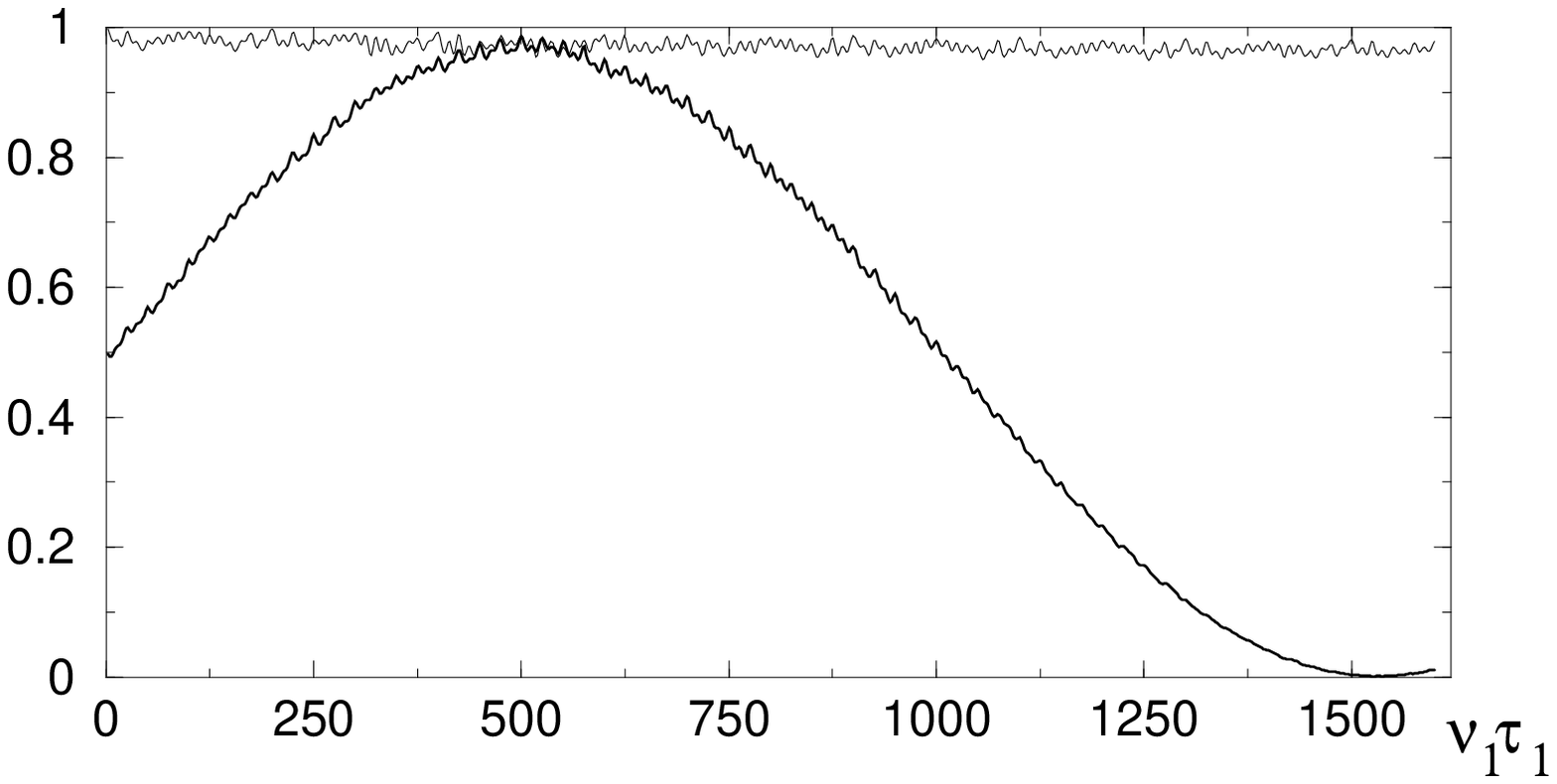}
\begin{minipage}{3.4truein}
\smallskip \smallskip \caption{Time evolution of ions in thermal motion.
In the bottom curve the initial internal state is
$\left|\psi_0\right\rangle=\left|+ -\right\rangle$, and we plot
the squared overlap over time with the Bell state
$\left|\beta_-\right\rangle$. In the top curve,
$\left|\psi_0\right\rangle=\left(\left|+
+\right\rangle+\left|--\right\rangle\right)/\sqrt{2}$, and we plot
the squared overlap over time with
$V^{\dag}(t)V(0)\left|\psi_0\right\rangle$. In both cases, the
motional modes are initially in thermal states with mean phonon
numbers $\bar{n}_1$ = 1, $\bar{n}_2$ = 0.1, undergoing heating at
rates $\Gamma_1 =
10^{-3}\nu_1$, $\Gamma_2 = 10^{-4}\nu_1$. Other parameter values
are $\eta_{11} = 0.025$, $\Omega = 1.5\nu_1$, $F = \nu_1/50$.}
\label{fig:evol}
\end{minipage}
\end{figure}
\end{center}

This robustness is illustrated in Fig. \ref{fig:evol}, where we
plot time evolution curves simulated using the full trap
Hamiltonian (i.e., including all orders of the LD parameter) and
allowing also for heating in the form of quantum jumps, described
by jump operators $\sqrt{\Gamma_p\bar{n}_p} a$ and $\sqrt{\Gamma_p
(\bar{n}_p+1)}a^{\dagger}$ \cite{Plenio98a}. (The heating rate
$\Gamma_1$ for the CM mode is assumed to be an order of magnitude
greater than $\Gamma_2$ \cite{James98b}). The curves are averages
obtained over $25$ Monte Carlo runs, each containing an average of
$18.4$ jumps. In the bottom curve, the ions' internal state is
initially $\left|\psi_0\right\rangle = \left|+ -\right\rangle$,
and we plot its squared overlap over time with the Bell state
$\left|\beta_-\right\rangle$. A fidelity of  $98\%$ is achieved
 at the time $\nu_1\tau _{1}\simeq$ 515
expected from Fig. \ref{fig:speed}. In the top curve,
$\left|\psi_0\right\rangle= \left(\left|+
+\right\rangle+\left|--\right\rangle\right)/\sqrt{2}$, and we plot
the squared overlap with
$V^{\dag}(t)V(0)\left|\psi_0\right\rangle$, the expected evolution
in the absence of the motion-state-dependent term in Eq.
(\ref{effHam}). The small deviation from 1 shows that this term
has been effectively suppressed by the  switching of the laser
phase.

At first sight, it may appear that ever greater suppression of the
motion-dependent evolution can be achieved by increasing the
switching rate $F$. One must be careful, however, not to
invalidate the approximations leading to Eq.~(\ref{effHam}). For
instance, $F$ must be limited by the Rabi frequency $\Omega$, in
order to allow the rotating-wave-type averaging implicit in the
derivation of Eq.~(\ref{JamesHam}) to hold within each interval
between phase inversions. In fact, our numerical simulations
indicate that this averaging only occurs fully for particular
"resonant" values of $F$. For most other choices, the time
evolution of the system gradually becomes degraded. Determining
analytically these `good' values is an open question; in practice,
one would likely ``tune" $F$ until an appropriate value was found.

In multi-ion chains with $N > 2$ our conclusions still hold, but
high-frequency phase shifts become a necessity even in the absence
of external heating. To see this, consider first that the LD
parameters $\eta _{jp}$ will generally differ from ion to ion
\cite{reviews}, and so $\left[ A_{p},B_{p}\right] \neq 0$. In this
case, inverting the sign of $B_{p} $ at a low frequency (e.g.,
comparable to $\omega $) is not sufficient to undo its effect on
states $\left| +-\right\rangle $ and $\left| -+\right\rangle $.
However, if the frequency is much larger than $\omega $ then this
cancellation does occur, since for very short timescales the
effect of the commutator above becomes negligible  \cite{Lloyd}. 
As a result, once again $B_{p}$ can be neglected, and the
Hamiltonian reduces effectively to the first term in Eq.
(\ref{effHam}). A further generalisation to the case where $m>2$
ions are illuminated together is also possible, as is the
counter-intuitive possibility of coupling {\em different} pairs of
ions simultaneously (cf the concluding remark in
\cite{Sorensen99a}). The latter case is achievable by illuminating
each pair with a separate beam of sufficiently different
intensity. In this case, following again the argument in Eqs.
(\ref{drespic})-(\ref{gatefreq}) shows that the Hamiltonian
effectively decouples into a sum of terms like in Eq.
(\ref{effHam}), one for each pair. Thus, adding a further
mind-boggling twist to the situation in \cite{Sorensen99a}, {\em
every single} vibrational mode can be used collectively and
simultaneously for different tasks, despite only ever being
`virtually' excited!

We thank A. S\o rensen, A. Steane and P. Knight for helpful
discussions. This work was supported in part by the Brazilian agency Conselho
Nacional de Desenvolvimento Cient\'{\i}fico e Tecnol\'{o}gico
(CNPq), EPSRC, the Leverhulme Trust, the European Science
Foundation's QIT programme, and the EQUIP project of the European
Union.

\end{multicols}


\begin{references}
\bibitem{Sornote} Note that, unlike in ref. \cite{Sorensen99a}, it
does not seem to be possible to achieve this without rapid
switching. In that case it turned out that a similar cancellation
could occur simply by shining both beams on both ions
simultaneously. Doing something similar here would result in no
effect, as the two phase-opposed beams would just cancel out!
\end{references}
\end{document}